\documentclass[letterpaper]{article} 
\usepackage{aaai25}  
\usepackage{times}  
\usepackage{helvet}  
\usepackage{courier}  
\usepackage[hyphens]{url}  
\usepackage{graphicx} 
\usepackage{natbib}  
\usepackage{caption} 
\usepackage{algorithm}
\usepackage{algorithmic}

\usepackage{microtype}
\usepackage{amsmath}
\usepackage{cleveref}
\usepackage{subfigure}
\usepackage{booktabs} 
\usepackage{multirow}

\usepackage{amssymb}
\usepackage{mathtools}
\usepackage{amsthm}
\usepackage{adjustbox}

\usepackage{newfloat}
\usepackage{listings}

\theoremstyle{plain}
\newtheorem{theorem}{Theorem}[section]

\newtheorem{prerequisite}[theorem]{Prerequisite}

\theoremstyle{definition}

\theoremstyle{remark}

\DeclareCaptionStyle{ruled}{labelfont=normalfont,labelsep=colon,strut=off} 
\floatstyle{ruled}
\newfloat{listing}{tb}{lst}{}
\floatname{listing}{Listing}
\title{Speech Watermarking with Discrete Intermediate Representations}
\author{
    Shengpeng Ji\equalcontrib\textsuperscript{\rm 1}, Ziyue Jiang\equalcontrib\textsuperscript{\rm 1}, Jialong Zuo\textsuperscript{\rm 1}, Minghui Fang\textsuperscript{\rm 1}, Yifu Chen\textsuperscript{\rm 1}, Tao Jin\textsuperscript{\rm 1}, Zhou Zhao\thanks{Corresponding author.}\textsuperscript{\rm 1}
}
\affiliations{
    \textsuperscript{\rm 1}Zhejiang University\\

    $\{$shengpengji, zhaozhou$\}$@zju.edu.cn
}

\usepackage{bibentry}

\begin{document}

\maketitle

\begin{abstract}
Speech watermarking techniques can proactively mitigate the potential harmful consequences of instant voice cloning techniques. These techniques involve the insertion of signals into speech that are imperceptible to humans but can be detected by algorithms. Previous approaches typically embed watermark messages into continuous space. However, intuitively, embedding watermark information into robust discrete latent space can significantly improve the robustness of watermarking systems. In this paper, we propose DiscreteWM, a novel speech watermarking framework that injects watermarks into the discrete intermediate representations of speech. Specifically, we map speech into discrete latent space with a vector-quantized autoencoder and inject watermarks by changing the modular arithmetic relation of discrete IDs. To ensure the imperceptibility of watermarks, we also propose a manipulator model to select the candidate tokens for watermark embedding. Experimental results demonstrate that our framework achieves state-of-the-art performance in robustness and imperceptibility, simultaneously. Moreover, our flexible frame-wise approach can serve as an efficient solution for both voice cloning detection and information hiding. Additionally, DiscreteWM can encode 1 to 150 bits of watermark information within a 1-second speech clip, indicating its encoding capacity. Audio samples are available at~\url{https://DiscreteWM.github.io/discrete_wm}.
\end{abstract}

\begin{figure}[ht]
\vskip 0.2in
\begin{center}
\centerline{\includegraphics[width=\columnwidth]{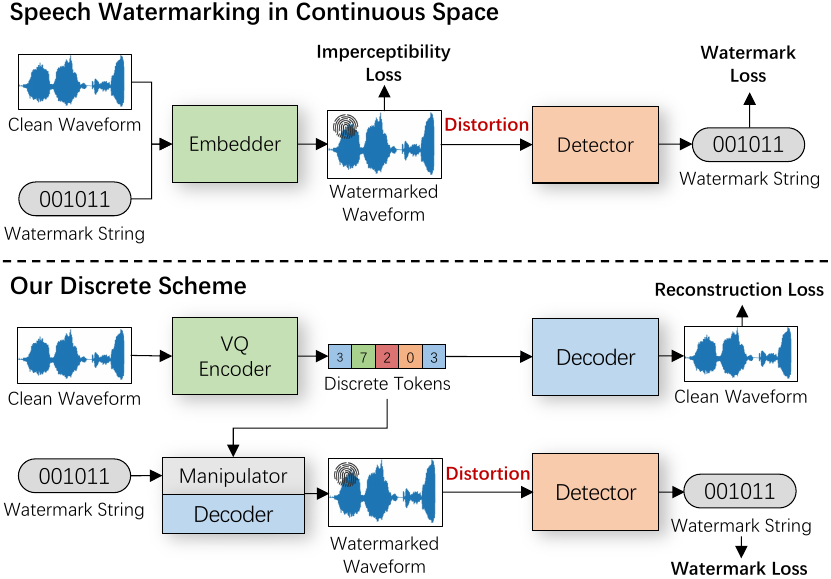}}
\caption{Illustration for speech watermarking strategies. \textbf{Upper: } The embedder learns to encode the watermark string into the continuous space with imperceptibility loss and watermark loss. \textbf{Lower: }In our discrete scheme, the vector-quantized variational autoencoder (VQVAE) maps speech into discrete latent space, and the manipulator conceals the watermark string within the modulus relations of discrete token IDs.}
\label{fig_1}
\end{center}
\vskip -0.2in
\end{figure}



\section{1\quad Introduction}
In recent years, the significant breakthrough in zero-shot text-to-speech (TTS)~\cite{casanova2022yourtts,wang2023neural,shen2023naturalspeech,le2023voicebox,jiang2023mega,mobilespeech,controlspeech,textrolspeech,funaudiollm} enables instant voice cloning with only a few seconds of speech. However, this technological advancement also brings security concerns to personal voices~\cite{seamless_communication,liu2023asvspoof}. To avoid potential misuse of voice cloning technology, passive detection strategies~\cite{tak2022automatic,ahmed2020void,tak2022rawboost,tak2021end} are developed to classify whether a speech clip is synthesized and adversarial-based methods~\cite{huang2021defending,li2023voice, languagecodec,yu2023antifake} are proposed to prevent voice cloning with adversarial noise. However, these approaches still struggle with generalization issues~\cite{liu2023detecting}. In comparison, speech watermarking has been developed to~\cite{pavlovic2022robust,liu2023dear,chen2023wavmark,liu2023detecting} proactively embed robust watermark information into the target voice, which has demonstrated its generalizable performance in practical voice cloning detection~\cite{seamless_communication}. By utilizing this technology, users can not only identify whether a speech clip is AI-generated but also trace the source of the speech, thus offering reliable privacy protection in the era of large-scale voice models.

\begin{figure*}[t]
\centering
  \subfigure[Ground Truth] {
    \includegraphics[width=0.47\columnwidth]{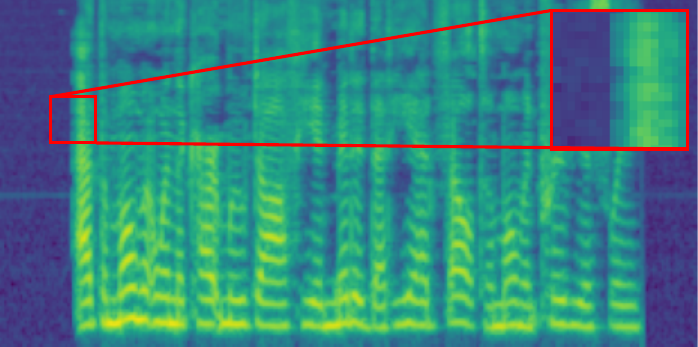}
    \label{fig:1}
  }
  \centering
  \subfigure[WavMark, capacity=32 bit] {
    \includegraphics[width=0.47\columnwidth]{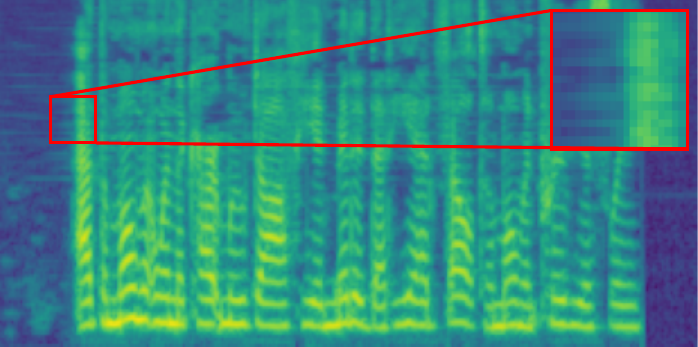}
    \label{fig:2}
  }
  \subfigure[Chang Liu's, capacity=30 bit] {
    \includegraphics[width=0.47\columnwidth]{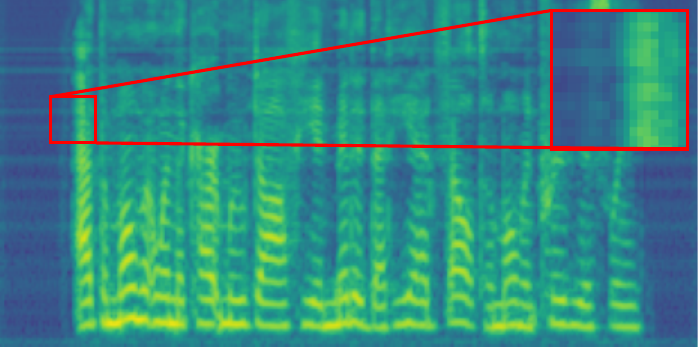}
    \label{fig:3}
  }
  \subfigure[Ours, capacity=32 bit] {
    \includegraphics[width=0.47\columnwidth]{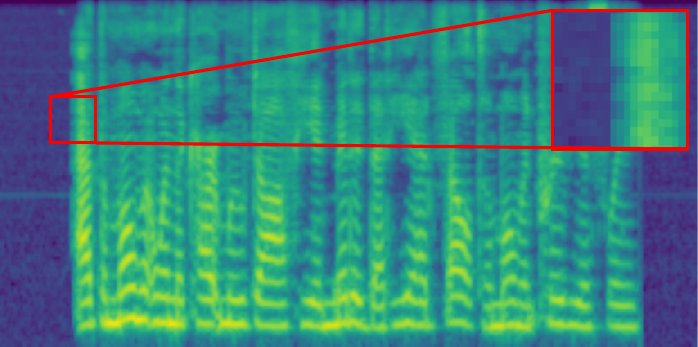}
    \label{fig:4}
  }
\caption{Visualizations of the ground-truth and watermarked mel-spectrograms by different speech watermarking methods. For a fair comparison, we directly download the example from WavMark's demo page and use the pre-trained Chang Liu's model.}
\label{vis_diff_method_1}
\end{figure*}

Despite recent advances in speech watermarking, current solutions still encounter two primary challenges: 1) trade-off among imperceptibility, robustness, and encoding capacity; In other words, maintaining robustness against various distortions while preserving a high encoding capacity affects the imperceptibility of watermarks~\cite{liu2023dear}. Although GAN-based architectures have been introduced to minimize the distribution difference between watermarked speech and clean speech, the embedder still encodes the watermark into perceptible noise patterns in the mel-spectrogram, as shown in Figure \ref{vis_diff_method_1}; 2) fixed length issues; Most DNN-based speech watermarking methods can only process a fixed length of waveform with a pre-defined length of watermark string. In the detection stage, they require a sliding window to decode a watermark starting at each frame~\cite{chen2023wavmark}, which is inefficient and constrains the resolution of watermarks to speeches larger than one second~\cite{seamless_communication}. Although some works integrate time-independent features into the watermarking algorithm~\cite{liu2023detecting}, the capacity of the watermark string can not be changed during the inference stage, which also limits the resolution of watermarks and affects the flexibility in handling various scenarios.

Intuitively, compared to encoding watermarks into continuous latent space, watermarks in robust discrete latent space are more robust against distortions. Therefore, to achieve a superior trade-off among imperceptibility, robustness, and encoding capacity, we propose DiscreteWM, a framework that utilizes discrete speech representations to embed watermark information. As shown in Figure \ref{fig_1}, we first propose a masked vector-quantized variational autoencoder (VQVAE) to map clean speech into frame-level discrete latent space. We ensure that the parity of the discrete token IDs can be detected from the reconstructed speech even when it is severely distorted. Then we propose a manipulator model to learn the probability distribution of discrete speech tokens. Finally, the watermark information can be embedded into the modular arithmetic relationship of discrete token IDs selected by the manipulator model. By utilizing the modular arithmetic relationship of discrete acoustic tokens in the latent space, our work enjoys an imperceptible and flexible watermarking pipeline where the users can freely decide the strength, capacity, and formats of the watermark information in the inference stage.

The contributions of the paper are summarized as follows:

\begin{itemize}
\item DiscreteWM is the first attempt to embed watermark information in the robust discrete latent space. Our method outperforms other state-of-the-art (SOTA) speech watermarking models on both voice cloning detection and information hiding tasks.

\item Our frame-wise strategy also resolves the challenges related to fixed-length training in speech watermarking and achieves 22.1x times faster detection speed.

\item DiscreteWM allows users to freely manipulate the encoding capacity (up to 150 bits per second) and formats of the watermark without re-training the model. 

\item We further propose a statistical Z-test to transform our frame-wise accuracy to utterance level for AI-generated content detection. The extensive studies demonstrate that our method achieves a false positive rate of $3 \times 10^{-5}$ while maintaining extreme imperceptibility.

\end{itemize}

\section{2 \quad Related Works}
\subsection{2.1 \quad Speech Watermarkiing}
Speech watermarking technology has always been used as a fundamental tool for copyright protection of human speech~\cite{hua2016twenty}. Traditional speech watermarking typically embeds watermark information in the time domain (e.g., Least Significant Bit~\cite{cvejic2004increasing}, Echo Hiding~\cite{gruhl1996echo}) and the transform domain (e.g., Spread Spectrum~\cite{cox1997secure}, Patchwork~\cite{yeo2003modified}). In terms of robustness, some researches have successfully achieved resilience against distortion~\cite{zhang2023m}, desynchronization~\cite{zhao2021desynchronization}, re-recording~\cite{liu2018patchwork}, etc. However, the encoding process of traditional methods relies heavily on hand-crafted empirical rules, which are challenging to implement, resulting in a low encoding capacity with limited robustness against a wider range of attacks.

Recently, DNN-based speech watermarking algorithms~\cite{jiang2020smartsteganogaphy,pavlovic2022robust,liu2023dear,chen2023wavmark,liu2023detecting,seamless_communication} have demonstrated superior encoding capacity, invisibility, and robustness when compared to traditional methods. Their frameworks typically include an encoder for watermark embedding and a detector for watermark extraction. The encoding and decoding strategies are learned in an end-to-end manner. In terms of imperceptibility, DeAR~\citep{liu2023dear} utilizes an adversarial discriminator to minimize the domain gap between clean speech and watermarked speech. WavMark~\cite{chen2023wavmark} regards the encoding and decoding as reciprocal processes and adopts invertible neural networks, which improves the overall fidelity and robustness of the watermark. And in terms of robustness, some of the most advanced methods can resist voice cloning attacks~\cite{liu2023detecting}, desynchronization attacks~\cite{chen2023wavmark}, and re-recording attacks~\cite{liu2023dear}. However, most of their methods, unfortunately, have limitations in that they can only process speech signals of a predetermined length. In order to locate the watermark, they rely on the Brute Force Detection (BFD) method, which involves sliding through the speech and attempting to decode a watermark starting at each frame~\cite{seamless_communication}. The latency of these approaches is excessively high, making them impractical as proactive defense mechanisms for real-world voice cloning systems. Besides, current solutions only embed the watermark as continuous noise patterns, leaving speech watermarking with discrete intermediate representation unexplored. Therefore, we propose a frame-wise approach to solve the watermark localization issues and investigate the algorithm that adopts discrete intermediate representations to further enhance the imperceptibility and robustness of watermarks. We include additional discussions about the vector quantised discrete representation and its applications in Appendix F.

\section{3 \quad Method}
\label{sec:method}
This section introduces DiscreteWM. To begin with, we provide an intuitive formulation and prerequisites of our watermarking strategy. Next, we provide detailed descriptions of our architecture design and the training process of the proposed model. Finally, we propose inference strategies for information hiding and AI-generated content detection separately and propose a statistical measure for detecting the watermark with the one proportion Z-test. Due to space
limitations, we provide technical details in Appendix A.

\begin{figure*}[!t]
    \centering	\includegraphics[width=0.99\textwidth]{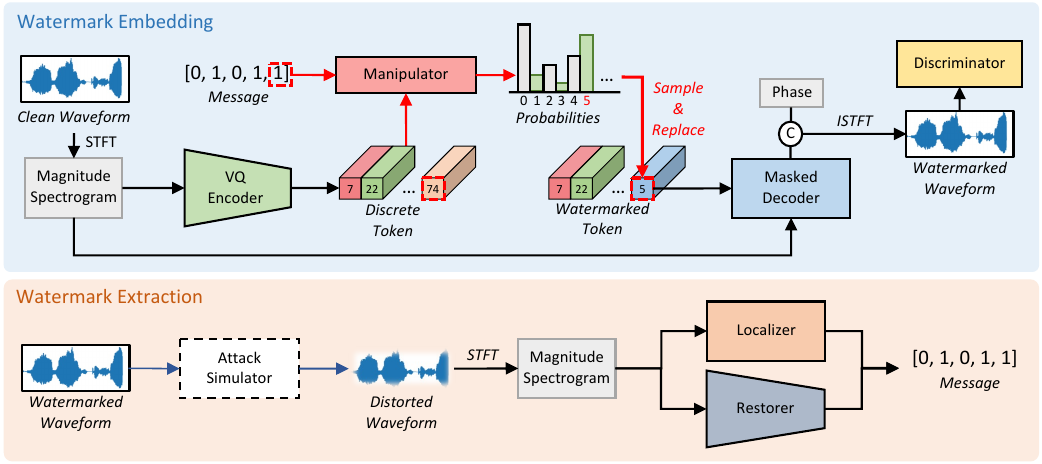}
	\caption{The overall architecture of DiscreteWM. ``VQ'' represents the ``vector quantization'' operation, and \textcircled{\tiny{C}} denotes the concatenation operation. During the \textit{watermark embedding} process, the manipulator forces the discrete tokens to have the same modular arithmetic relation with the watermark message, as indicated by the red dashed line. For instance, if we intend to conceal the value ``1'' into the last discrete token, the manipulator will selectively sample from the odd tokens (highlighted in green) according to their probability distribution. The original token will then be replaced with the sampled token that has the highest probability (the 5th token). In \textit{watermark extraction}, the localizer is responsible for watermark localization, while the restorer focuses on recovering the watermark message.}
	\label{fig:arch_overview}
\end{figure*}

\subsection{3.1 \quad Watermarking Strategy}
\label{sec:watermarking_strategy}
The outline of our watermark strategy is: \textit{transforming speech into discrete latent space and enforcing the discrete token IDs to have the same modular arithmetic relations with the watermarks}. \\
\textbf{Strategy formulation.}  
Denote $s=\{s^{(0)}, \cdots, s^{(T)}\}$ as the magnitude spectrogram of speech waveform $y$ and $w$ as the watermark string, where $T$ is the number of spectrogram frames. The watermark embedding process is performed according to the following steps: 1) an encoder $\mathbf{E}$ learns to represent the spectrogram $s$ with acoustic code sequence $z=\{z^{(0)}, \cdots, z^{(T)}\}$, where $z^{(t)}$ is obtained from a discrete codebook $\mathcal{Z}$; 2) Then, we inject the watermark string $w$ into $z$ by manipulating the modulus relation of token IDs $c$. For simplicity, we only consider the case of ``$c \bmod 2$'' in this section, as it is a suitable setting for speech watermarking~\cite{chen2023wavmark}. Specifically, when we want to embed the watermark character ``0'' or ``1'' in the $t$-th frame, we replace the $t$-th discrete code with the even or odd code ID that has features similar to the original one, respectively. The watermarked acoustic codes are denoted as $\hat{z}$; 3) Given $\hat{z}$, a decoder $\mathbf{G}$ learns to reconstruct the watermarked spectrogram $\hat{s}$. $\hat{s}$ and the original phase spectrogram are converted to the watermarked speech $\hat{y}$ through the inverse Short-Time Fourier Transform operation (iSTFT); 4) A localizer $\mathbf{D}$ is designed to locate the watermarked frames and a restorer $\mathbf{R}$ is utilized to recover $\hat{z}$. Finally, we can obtain the watermark string $w$ from $\hat{z}$.\\
\textbf{Prerequisites of the proposed strategy.} 
However, the above strategy can not guarantee the imperceptibility and robustness of the watermark until now. In practical scenarios, in terms of imperceptibility, the perceptual differences of $y$ and $\hat{y}$ should be minimized. Therefore, the proposed watermarking strategy needs the following prerequisites: 
\begin{prerequisite}
\label{eq:prerequisite_1}
$\mathbf{G}\left(z\right) = \bar{s} \to s$, the difference between the reconstructed spectrogram $\bar{s}$ and the original spectrogram $s$ should be minimized.
\end{prerequisite}
\begin{prerequisite}
\label{eq:prerequisite_2}
$\hat{z} \to z$, the distance between the manipulated acoustic code $\hat{z}$ and the original code $z$ in the latent space should be minimized.
\end{prerequisite}
\noindent
In terms of robustness, it is crucial to accurately extract the watermark string $w$ even when $\hat{y}$ is distorted in signal transmission processes or is maliciously attacked:
\begin{prerequisite}
\label{eq:prerequisite_3}
$\mathbf{R}\left(\mathbf{D}\left(Dist\left(\hat{y}\right)\right)\right) \to \hat{c} \bmod 2 = w$, where $Dist\left(\cdot\right)$ is the distortion function. 
\end{prerequisite}
We describe how we achieved the aforementioned prerequisites in the following subsection.


\subsection{3.2 \quad Architecture Design}
\label{sec:atchitecture_design}
Our framework comprises a two-stage training process. \textit{In the first stage}, we train an autoencoder to represent the speech into discrete tokens. Then, we construct a localizer model $D$ to locate the reconstructed frames and design a restoration loss to ensure $\mathbf{R}$ can restore the parity of discrete token IDs ($\hat{c} \bmod 2$) even when the reconstructed speech is heavily distorted. \textit{In the second stage}, we train a probability-based manipulator model to conceal the watermark string within the modular arithmetic relationships among these discrete tokens while ensuring imperceptibility.

\subsubsection{3.2.1 \quad Robust Discrete Latent Space\\}
\label{sec:robust_discrete_latent_space}

\textbf{Representing speech in discrete latent space.}  
Given a clean speech $y$, we first represent it in the discrete latent space. As shown in Figure~\ref{fig:arch_overview}, we apply the Short-Time Fourier Transform operation (STFT) on $y$ to produce a magnitude spectrogram $s$. Then, to discretize $s$, we adopt a vector quantized variational autoencoder architecture (VQVAE)~\cite{van2017neural}. The VQ encoder $\mathbf{E}$ and decoder $\mathbf{G}$ reconstruct the spectrogram $s$ through: $\bar{s}=\mathbf{G}\left(z\right)=\mathbf{G}\left(\mathbf{E}\left(s\right)\right)$. Additionally, to satisfy Prerequisite \ref{eq:prerequisite_1}, the system is trained through a mask-infilling process with a frame-level random mask. Due to the spectro-temporal locality of speech signals~\cite{espi2015exploiting}, the unmasked contextual speech can provide rich information to significantly reduce the difficulty of the spectrogram reconstruction. The discrete codes of the masked region are also fed into the decoder to provide the missing information during the masking process. Finally, the reconstructed spectrogram of the masked region is concatenated with the unmasked original spectrogram. The overall reconstruction process $\bar{s} \approx s$ is formulated as:

\begin{align} 
\label{eq:eq_1}
\bar{s}=\omega \cdot \mathbf{G}\left(\omega \cdot \mathbf{E}\left(s\right) , \left(1-\omega\right) \cdot s \right) + \left(1-\omega\right) \cdot s\ , 
\end{align}
where $\omega$ is the binary mask. $\omega$ is obtained by $\omega = \textit{Mask}\left(s, \gamma\right)$, where $\textit{Mask}\left(\cdot\right)$ is the mask function and $\gamma \in [0.1, 0.5]$ is the mask ratio. To further minimize the perceptual differences between $\hat{y}$ and $y$, we introduce extra discriminators for adversarial training, including the multi-period discriminator and the multi-scale discriminator~\cite{kong2020hifi}. Finally, the training loss of the VQ-VAE can be formulated as:
\begin{align} 
\label{eq:eq_2}
\mathcal{L_{VQ}} = \mathcal{L}_{\mathrm{rec}} + \mathcal{L}_{\mathrm{code}} + \lambda_{adv}\mathcal{L}_{\mathrm{adv}}\ ,
\end{align}
where $\mathcal{L}_{\mathrm{rec}}$ is the reconstruction loss, $\mathcal{L}_{\mathrm{code}}$ is the standard VQ codebook loss~\citep{van2017neural}, and $\mathcal{L}_{\mathrm{Adv}}$ is the adversarial loss. We use the multi-resolution STFT loss~\cite{yamamoto2020parallel} as $\mathcal{L}_{\mathrm{rec}}$. $\lambda_{adv}$ is the hyper-parameter to balance the three terms, which is set to $10^{-2}$. To enhance the codebook usage rate and further decrease the reconstruction error, we adopt the clustering vector quantizer (CVQ)~\cite{zheng2023online} as the element-wise quantization function in $E$ that maps each acoustic code onto its closest codebook entry.\\ 
\textbf{Detecting the Parity of Token IDs.} 
Here we describe how to restore the parity of discrete token IDs ($\hat{c} \bmod 2$) from the reconstructed speech, which is the necessary condition for watermark embedding in the discrete latent space. As shown in Figure~\ref{fig:arch_overview}, our frame-wise framework has two primary objectives: \textit{localization} and \textit{discrete code restoration}. Regarding \textit{localization}, we aim at distinguishing between the original frames and the reconstructed frames with the localizer model $\mathbf{D}$; We train $\mathbf{D}$ by minimizing the binary cross-entropy loss between its output and a binary mask representing the presence of the reconstructed frames. With the localizer model $\mathbf{D}$, our algorithm successfully resolves the location issues in current fixed-length counterparts. Compared to the previous sliding-window detection method, the proposed localizer significantly reduces the time required for watermark localization. In terms of \textit{discrete code restoration}, we focus on converting the reconstructed speech $\hat{y}$ back to the manipulated discrete token $\hat{z}$ using the restorer model $\mathbf{R}$ even when $\hat{y}$ is severely distorted. We design the following restoration loss to achieve this objective:
\begin{align} 
\label{eq:eq_3}
\mathcal{L}_{res} = \mathbb{E}_{\tilde{s} \sim p(\tilde{s})}[-\log p(\hat{c} \bmod 2)],
\end{align} 
where $\tilde{s}$ is the magnitude spectrogram of $Dist(\hat{y})$ and $\hat{c}$ is the token IDs of $\hat{z}$. Furthermore, to fulfill Prerequisite~\ref{eq:prerequisite_3}, an attack simulator is employed in our framework following previous works~\cite{chen2023wavmark,liu2023detecting}, which assists our model in acquiring adaptive robustness against various attacks $Dist\left(\cdot\right)$. Until now, we have finally built a robust discrete latent space, in which the parity of the discrete code IDs can be easily detected.

\subsubsection{3.2.2 \quad Injecting Watermarks into Discrete Latent Space\\}
\label{sec:injecting_watermarks_into_discrete_latent_space}
\textbf{Concealing watermarks with the manipulator.}      
As illustrated in Section 3.1, our DiscreteWM embeds watermarks by ensuring that the discrete token IDs have identical modular arithmetic relationships with the watermarks. However, if we manually adjust the code IDs to embed watermark information, it will have a significant impact on the speech quality. For instance, if we replace the discrete code representing silence with the discrete code of normal speech, there will be a significant amount of noise in the watermarked frame. To satisfy Prerequisite~\ref{eq:prerequisite_2}, we introduce a probability-based manipulator model $\mathbf{M}$ to help us select the optimal code ID in the watermark embedding process. During the second-stage training process, we first extract $z$ through $\mathbf{E}\left(s\right)$ using the proposed VQVAE structure. Given $\omega \cdot z$ as the prediction target, the manipulator model $\mathbf{M}$ is trained through a parallel mask-prediction process:
\begin{align}
\label{eq:eq_4}
P\left(\omega \cdot z \mid \left(1-\omega\right) \cdot z; \theta_{M}\right)\ ,
\end{align}
where $\omega$ is the aforementioned binary mask and $\theta_{\mathbf{M}}$ is the parameter of $\mathbf{M}$. The manipulator model is trained with the cross-entropy loss. After training, $\mathbf{M}$ can be utilized to sample the odd or even optimal tokens according to the watermark information and replace the original discrete token to construct $\hat{z}$. \\
\textbf{Sampling strategy of the manipulator.}  
As shown in Figure~\ref{fig:arch_overview}, to embed the watermark value ``1'' into the last frame, if the ID value of the last discrete token is even, we replace it with odd tokens sampled from the probability distribution given by the manipulator model $\mathbf{M}$:
\begin{align}
\label{eq:eq_5}
P(z^{(t)}_{k}) = softmax(l^{(t)}_{k}) = \frac{\mathrm{e}^{l^{(t)}_{k}}}{\sum_{i} \mathrm{e}^{l^{(t)}_{i}}}, 
\end{align}
where $l^{(t)}_{k}$ represents the logit of token $k$ at timestep $t$. If the ID value of the last discrete token is odd, we directly use the original token for reconstruction. During the watermark embedding process, we randomly select a portion of the discrete codes and substitute them non-autoregressively to ensure the efficiency of the system.

\subsection{3.3 \quad Inference Strategies}
During the inference stage, our frame-wise solution offers remarkable flexibility, enabling us to select different encoding strategies for various scenarios and to freely control the trade-off between imperceptibility and robustness. In this subsection, we discuss the watermark strategies for \textit{information hiding} and \textit{AI-generated content detection} separately. Additionally, we perform a statistical analysis on the detection sensitivity of the watermarked speech using the one proportion Z-test.

\subsubsection{Watermark for Information Hiding.}
Speech watermarking for information hiding mainly aims at hiding a binary message (such as 32 bits) to the speech segments~\cite{liu2023detecting,chen2023wavmark}, which can be used for tracing provenance, copyright protection, and privacy protection. The basic idea of our frame-wise watermarking strategy, as mentioned in Section 3.1, is to embed the
watermark character “0” or “1” by enforcing the token ID to be even or odd, respectively. In the information hiding pipeline, we first map clean speech into discrete latent space following Section 3.2.1 and embed watermark information into the discrete codes following Section 3.2.2. Then, the watermarked latent codes $\hat{z}$ are converted into the watermarked speech $\hat{y}$. Finally, following the watermark detection algorithms described in Section 3.2, we can recover the watermark string from $\hat{y}$. Since our watermarking method is frame-wise, it is free from the time-consuming watermark localization process like previous DNN-based methods~\cite{chen2023wavmark}. Moreover, our framework can freely adjust the encoding capacity according to users' requirements. Suppose the hop size is set to 80 and the maximum mask ratio $\gamma$ is set to 50\%, we can store 1 to 150 bits of information within one-second speech sampled at 24 kHz, which demonstrates the flexibility of our method.

\subsubsection{Watermark for AI-Generated Detection.}\label{method:wm_ai_detection}Speech watermarking is a crucial proactive defense strategy against voice cloning attacks~\cite{seamless_communication}. In this scenario, online services or individual users can add watermarks when cloning voices. In this way, people can easily determine whether the speech is generated by AI through the watermark detection process, which significantly reduces the possible abuses of voice cloning techniques. 

As discussed in Section 3.2, our localizer $\mathbf{D}$ can be employed to identify whether a speech frame is reconstructed by our VQVAE or not. Therefore, we can utilize this characteristic to achieve AI-generated content detection. In an ideal scenario, when a natural speech is given as input, the localizer $\mathbf{D}$ should output a sequence of zeros. If any frame in the output sequence of $\mathbf{D}$ is non-zero, it indicates that the audio segment has been watermarked, i.e., the audio segment is generated by AI. However, in practical situations, the frame-wise accuracy of $\mathbf{D}$ will ultimately affect our decision. In order to \textit{convert the frame-wise accuracy to utterance level}, we adopt a Z-test as our robust detection approach. In practical scenarios, we can detect the utterance-level watermark if the Z-statistic is above a pre-defined threshold (e.g., Z-statistic $> 4$). Denote $T$ as the number of speech frames. Let's assume that the frame-level true positive rate and false positives rate of $\mathbf{D}$ on the test set are $\alpha$ and $\beta$, respectively. Then, given a clean speech $y$, the number of its detected watermarked frames $|f|_{w}$ has expected value $\beta \cdot T$ and variance $\beta \left(1-\beta \right) \cdot T$. The Z-statistic can be calculated as:
\begin{align} 
\label{eq:eq_6}
\textit{Z-statistic} = \frac{\left(|f|_{w} - \beta \cdot T \right)}{\sqrt{\beta\left(1-\beta\right) \cdot T}}.
\end{align}
Denote $m=10\%$ as the watermark ratio and let $\alpha=95\%$, $\beta=10\%$, and $T=200$. In the detection stage, a watermarked speech will produce $|f|_{w} = \alpha \cdot m \cdot T + \beta \cdot (1-m) \cdot T = 37$, which means the z-statistic is $4.01$ and the one-sided p-value is $3\times10^{-5}$ approximately. In this case, the utterance-level probability of a false positive is only $3\times10^{-5}$, indicating that the watermark can be easily detected with extremely high confidence. Moreover, since $m$ can be adjusted in inference, users are free to decide whether to add more watermarks to enhance robustness or reduce watermarks to enhance imperceptibility. The summary of the proposed inference strategies is in Appendix D.

\begin{table*}[t]\small
\caption{Comparison with existing speech watermarking methods for information hiding. ``MEAN'' represents the average BER. ``Ours-\textit{32bps}'' means we insert 32 bits of watermark information to one-second speech segments in inference.}
\label{table:results_of_information_hiding}
\vskip 0.15in
\begin{center}
\begin{tabular}{l|ccc|ccccccccc}
\toprule
\multirow{2}{*}{Models} & \multirow{2}{*}{BPS($\uparrow$)} & \multirow{2}{*}{PESQ($\uparrow$)} & \multirow{2}{*}{SNR($\uparrow$)} & \multicolumn{9}{c}{BER(\%)($\downarrow$)} \\
             &    &      &       & ND   & GN   & AS   & RS   & MP3  & MF   & LP & EA & MEAN \\
\midrule
Audiowmark$^{*}$ & 20 & 4.39 & 29.85 & 5.89 & 18.13& 5.89 & 15.10 & 6.65 & 12.83 & 5.89 & 7.61 & 9.75 \\
DeAR         & 8.8& 3.75 & 26.31 & 0.45 & 0.48 & 0.46 & 0.42 & 0.58 & 0.48 & 0.91 & 0.51 & 0.54\\
Chang Liu's  & 30 & 3.97 & 24.18 & 0.00 & 2.68 & 0.00 & 0.02 & 0.00 & 0.06 & 0.00 & 0.04 & 0.35\\
WavMark      & 32 & 4.31 & \textbf{38.61} & 0.43 & 5.72 & 0.61 & 0.65 & 0.56 & 6.07 & 2.08 & 4.49 & 2.58\\
\midrule
Ours-\textit{32bps}   & \textbf{32} & \textbf{4.45} & 38.08 & 0.12 & 0.73 & 0.12 & 0.17 & 0.13 & 0.69 & 0.19 & 0.12 & \textbf{0.28}\\
\bottomrule
\end{tabular}
\end{center}
\vskip -0.1in
\end{table*}

\section{4\quad Experiments} 
\label{sec:Experiments}

\subsection{4.1\quad Experimental Setup}
\label{sec:Experimental_Setup}
\textbf{Datasets.}  
\textit{For training}, we employ the standard training set of LibriTTS~\cite{zen2019libritts}, which contains approximately 585 hours of English speech at 24kHz sampling rate. For the Short-Time Fourier Transform operation (STFT), we adopt a filter length of 400, a hop length of 80, and a window function applied to each frame with a length of 400. In our experiment, we find that a smaller hop length will increase the encoding capacity of the watermark, but setting the hop size too small is harmful for speech reconstruction. \textit{For evaluation}, we adopt two state-of-the-art zero-shot voice cloning models, NaturalSpeech 2~\cite{shen2023naturalspeech} and Mega-TTS 2~\cite{jiang2023mega2}, to generate high-quality synthesized audio that sounds authentic. We randomly select 100 text transcriptions and 100 speech prompts from the LibriTTS test-clean set. Each speech prompt is fed into the voice cloning model to generate speeches according to the 100 target sentences. The test set also includes all of the speech samples from the ``test-clean'' set of LibriTTS. As a result, a test set consisting of 24,837 sentences is obtained, with all speakers in the test set being unseen. We use all samples in the test set for evaluations. We provide implementation details in Appendix A.
\textbf{Evaluation Metrics.}  
For \textit{imperceptibility}, we adopt Signal-to-Noise Ratio (SNR) and Perceptual Evaluation of Speech Quality (PESQ)~\cite{rix2001perceptual} as metrics following previous works~\cite{liu2023detecting}. Among them, SNR is only used to measure the magnitude of differences between the watermarked speech and the original speech.
In comparison, PESQ provides a more accurate assessment of imperceptibility by considering the specifics of the human auditory system. For evaluating the \textit{effectiveness and robustness} of watermark extraction, we use the bit error rate (BER) as the metric. For \textit{encoding capacity}, we use bit per second (BPS) as the metric, which refers to how many bits of watermark information can be injected into one second of speech.

\subsection{4.2\quad Results of Information Hiding}
\label{exp:results_of_information_hiding}
In this subsection, we compare our DiscreteWM with different baseline systems to evaluate its ability of information hiding. To demonstrate the performance of different models in a concise and fair manner, we conduct a segment-based evaluation where we randomly extract a 1-second speech segment from each test sample. In this evaluation, the models aim to watermark one-second speech clips while remaining robust against various distortions and maintaining imperceptibility. The distortions include: 1) no distortion (ND); 2) Gaussian noise (GN); 3) amplitude scaling (AS);  4) re-sampling (RS); 5) MP3 compression (MP3); 6) median filter (MF); 7) low-pass filter (LP); 8) echo addition (EA); We provide further explanation for these distortions in Appendix B. 

We compare our model with existing state-of-the-art (SOTA) neural network based methods: 1) Audiowmark~\cite{westerfeld2020audiowmark}, a SOTA traditional watermarking toolkit that utilizes the patchwork-based watermarking method~\cite{liu2018patchwork} and incorporates BCH codes~\cite{bose1960class} for error correction. We used the default setting of Audiowmark; 2) DeAR~\cite{liu2023dear}, one of the pioneer deep learning frameworks for robust speech watermarking; 3) Chang Liu's method~\cite{liu2023detecting}, a strong and robust baseline that embeds the watermark into the frequency domain; 4) WavMark~\cite{chen2023wavmark}, a concurrent solution that employs invertible neural networks (INN) to ensure the inaudibility and robustness. Since we found that Audiowmark can hardly embed watermarks into the one-second speech segment, we use the utterance-level evaluation for it. The encoding capacity of Audiowmark is referenced from previous works~\cite{chen2023wavmark}. Although WavMark has an encoding capacity of \textit{32bps}, it still requires 10 to 16 bits of information for watermark localization. 

Increased signal-to-noise ratio (SNR) and perceptual evaluation of speech quality (PESQ) score indicate higher imperceptibility, while a lower bit error rate (BER) represents superior robustness. In the imperceptibility evaluation, distortions are not applied to the watermarked speech. As shown in Table~\ref{table:results_of_information_hiding}, our speech watermarking method, referred to as ours-\textit{32bps}, achieves comparable imperceptibility to WavMark and is on par with Chang Liu's approach in terms of robustness. This indicates that our method achieves a superior balance between imperceptibility and robustness, thus further validating the effectiveness of the discrete representations.

\begin{table}[t]\small
\caption{Evaluation for AI-generated speech detection. ``MEAN'' represents the average BER across all distortions. The RTF (Real-Time Factor) evaluation is conducted with 1 NVIDIA A100 GPU and batch size 1.}
\label{table:results_for_AIdetection}
\vskip 0.15in
\begin{center}
\begin{tabular}{l|cc|c|c}
\toprule
Models & PESQ($\uparrow$) & SNR($\uparrow$) & MEAN($\downarrow$) & RTF($\downarrow$) \\
\midrule
WavMark      & 4.24 & 37.92 & 1.02 & 0.1438\\
SeamlessWM$^{*}$   & 3.77 & 29.62 & \textbf{0.18} & 0.0065 \\
\midrule
Ours    & \textbf{4.37} & \textbf{38.01} & 0.32 & \textbf{0.0044}\\
\bottomrule
\end{tabular}
\end{center}
\vskip -0.1in
\end{table}

\subsection{4.3 \quad Watermarking for AI-Generated Speech Detection}
\label{exp:results_of_ai_generated_speech_detection}
In this subsection, we compare our DiscreteWM with different baseline systems to evaluate its ability to effectively put and detect an imperceptible watermark on top of AI-generated speech. To ensure reliable protection across various audio lengths in real-world applications, it is important for the model to accurately locate the positions of the watermarks and decode the original watermark.  Therefore, in this experiment, we conduct an utterance-level evaluation. As for the baseline systems, in addition to Audiowmark and WavMark, we also include SeamlessWM~\cite{seamless_communication}, which is a state-of-the-art concurrent work focused on detecting AI-generated content. Since SeamlessWM does not provide the pre-trained models and source code, we use the reproduced version for our experiments. We evaluate the imperceptibility (PESQ and SNR), robustness (MEAN: the averaged BER (\%) across all distortions), and inference efficiency (RTF) of these systems. The distortions follow the same setting in Section 4.2. In addition, when measuring RTF, we include both the watermark embedding and detection processes. We set the watermark ratio $m$ of DiscreteWM to $10\%$.

The results presented in Table~\ref{table:results_for_AIdetection} indicate that our method achieves comparable robustness compared to SeamlessWM, while also exhibiting superior imperceptibility. It also demonstrates that our method can provide a highly effective and reliable security guarantee for online speech synthesis services. In terms of the inference speed, the RTF of WavMark is significantly higher than other methods. In the experiments, we find that the sliding window localization process costs most of its inference time. Meanwhile, compared with WavMark, our frame-wise solution speeds up the speech watermarking process by 22.1x.

\subsection{4.4\quad Ablation Studies}
\begin{figure}[ht]
\vskip 0.2in
\begin{center}
\centerline{\includegraphics[width=\columnwidth]{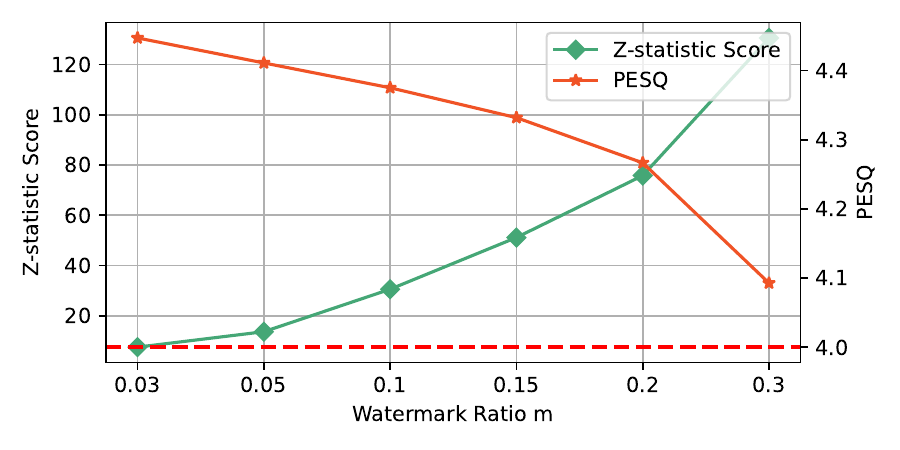}}
\caption{The tradeoff between reliability and imperceptibility on the AI-generated content detection task. ``Z-statistic = 4.0'' is shown as the red dashed line.}
\label{fig:vis_z_score}
\end{center}
\vskip -0.2in
\end{figure}

\textbf{Encoding Capacity.}
Our method can flexibly change the encoding capacity during the inference process. In this experiment, we evaluate the performance of DiscreteWM using various encoding capacities on the information hiding task. As shown in Table~\ref{table:ablation}, we can see that DiscreteWM maintains a high level of imperceptibility when its encoding capacity ranges from \textit{10} to \textit{50bps}, and it also performs well even under the extreme condition of \textit{150bps}. Additionally, the robustness of our method remains consistently high across different encoding capacities.\\
\textbf{Discrete vs Continuous.}
We evaluate the performance of DiscreteWM using discrete intermediate representation and continuous representation on the information hiding task. To make fair comparisons, we only remove the VQ layer and replace the manipulator with a watermark encoder to build the continuous baseline. The encoding capacity of the continuous baseline is set to \textit{32bps}. From Table~\ref{table:ablation}, it can be seen that our method with discrete intermediate representation achieved a better balance between imperceptibility and robustness than the continuous baseline, demonstrating the advantages of discrete intermediate representation.\\
\textbf{Manipulator vs Manual.}  
We test the effectiveness of the proposed manipulator model on the information hiding task. The encoding capacities of baseline systems in this experiment are set to \textit{32bps}. For ``wo/ manipulator'', we manually choose random codes for watermark embedding. The results in Table~\ref{table:ablation} demonstrate that without the manipulator, the imperceptibility of our method significantly drops, indicating the advantages of the proposed manipulator. \\
\textbf{Utterance-level Reliability.}  
In this experiment, we evaluated the utterance-level reliability of DiscreteWM on the AI-generated content detection task with the Z-test. The segment-wise methods like WavMark can only determine that the speech contains watermarks when the extracted watermark is the same as the preset one, which is not suitable for the proposed Z-test. Therefore, we do not compare our method with them here. In this evaluation, the watermarked speech is randomly attacked with the distortions following Section 4.2. We visualize the Z-statistic score (reliability) and PESQ (Imperceptibility) with different watermark ratios $m$ in Figure~\ref{fig:vis_z_score}. When the watermark ratio $m$ is $0.03$, the Z-statistic is 4.07. In this case, the false positive rate is only $2.3 \times 10^{-5}$. Moreover, given the Z-statistic=4.0 as the classification threshold, the utterance-level true positive rate and false positive rate are 1.0 and 0.0 when the watermark ratio is above 0.10. These results indicate that our method exhibits high imperceptibility while maintaining a high level of accuracy.

\begin{table}[t]\small
\caption{Ablation studies of DiscreteWM for information hiding.}
\label{table:ablation}
\vskip 0.15in
\begin{center}
\begin{tabular}{l|cc|c}
\toprule
Setting & PESQ($\uparrow$) & SNR($\uparrow$) & MEAN($\downarrow$) \\
\midrule
Ours-\textit{10bps}   & 4.47 & 41.30  & 0.31\\
Ours-\textit{32bps}    & 4.45 & 38.08 & 0.28\\
Ours-\textit{50bps}    & 4.27 & 34.92 & 0.30\\
Ours-\textit{150bps}   & 3.92 & 28.49 & 0.27\\
\midrule
w/ continuous    & 4.32 & 34.90 & 2.39 \\
\midrule
wo/ manipulator       & 3.96 & 29.55 & 0.95 \\
\bottomrule
\end{tabular}
\end{center}
\vskip -0.1in
\end{table}

\section{5\quad Conclusions}
In this paper, we present DiscreteWM, a framework that injects watermarks within the discrete intermediate representations of speech. Our approach outperforms the continuous counterparts in terms of robustness and imperceptibility. Besides, our frame-wise solution allows for encoding 1 to 150 bits of watermark information into only a 1-second speech clip, demonstrating its flexibility and encoding capacity. Moreover, the proposed utterance-level Z-test also indicates the reliability of our method for voice cloning detection.

\appendix

\section{Acknowledgments}
This work was supported by the National Natural Science Foundation of China under Grant No.62222211 and No.U24A20326

\bibliography{aaai25}

\section{A\quad Model and Training Details}
\label{app:model_configuration_training_details}

\subsection{A.1 Network Structure}
\begin{figure*}[!ht]
	\centering
	\includegraphics[width=0.75\textwidth]{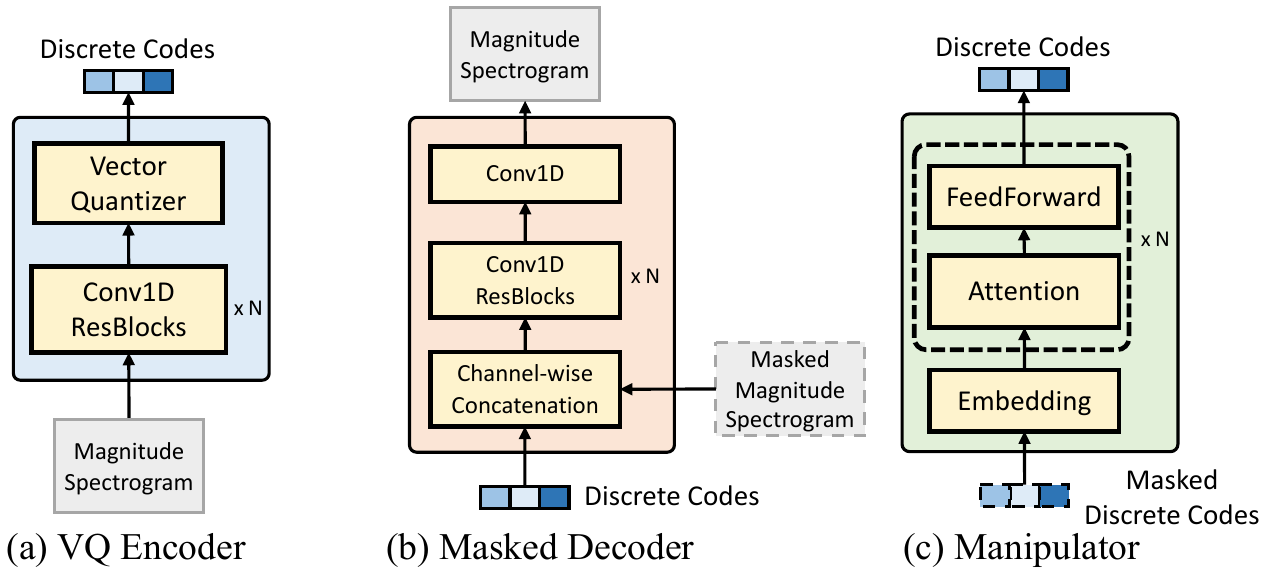}
	\caption{The structure of the VQ encoder, the masked decoder, and the manipulator.}
	\label{fig:detailed_arch}
\end{figure*}

\paragraph{VQ Encoder.} We visualize the network structure of the VQ encoder in Figure~\ref{fig:detailed_arch} (a). The VQ encoder maps the magnitude spectrogram into discrete codes with the convolutional residual blocks and the vector quantizer. The convolutional residual blocks consist of 3 1D convolutional blocks with 128 hidden size and 3 kernel size. We do not use pooling layers so that the information will pass through the VQ encoder as much as possible to minimize the reconstruction error.

\paragraph{Masked Decoder.} The detailed network structure of the proposed masked decoder is shown in Figure~\ref{fig:detailed_arch} (b), which utilized the discrete codes and masked magnitude spectrogram to reconstruct the original magnitude spectrogram. We first concatenate the discrete code embedding and the masked magnitude spectrogram in a channel-wise manner. Then, the features are fed into several 1D convolutional residual blocks. Finally, we use 1D convolution layer to map the output of the model to the magnitude spectrogram. The convolutional residual blocks consist of 3 1D convolutional blocks with 128 hidden size and 3 kernel size.

\paragraph{Manipulator.} As shown in Figure~\ref{fig:detailed_arch} (c), the manipulator is built with a stack of Transformer blocks~\cite{vaswani2017attention}, which aims at predicting the discrete code sequence given by the pre-trained VQ-VAE model in a non-autoregressive manner. The Transformer blocks consist of 4 Transformer layers with 128 hidden size and 2 attention heads.

\paragraph{Localizer and Restorer.} The localizer and restorer share the same architecture with the masked decoder. The input of the localizer and restorer is both the magnitude spectrogram. The localizer aims at locating the watermarked frames and the restorer recovers the watermark information from the located frames.

\paragraph{Codebook.} To solve the codebook collapse issue of the vanilla VQ-VAE~\citep{takida2022sq} and enhance the convergence of the training process, we adopt a dynamical initialization strategy based on CVQ-VAE~\citep{zheng2023online} during training, which ensures the code vectors that are less-used or unused to be modified more than frequently used ones. But we do not use the contrastive loss in CVQ-VAE to encourage code sparsity, which will affect the performance of our watermark detection. The codebook embedding size is 128 and the hidden size of the codebook vector is 128. 

\paragraph{Discriminator.} The discriminator follows the default architecture of the multi-period discriminators and multi-scale discriminator proposed in~\citet{kong2020hifi}.

\subsection{A.2 Model Configuration}
\label{app:model_configuration}
We provide the hyper-parameter settings of our DiscreteWM in Table~\ref{table:model_configuration}.

\begin{table}[h]
\caption{Model configuration of DiscreteWM.}
\label{table:model_configuration}
\centering
\begin{tabular}{@{}llcc@{}}
\toprule
\multicolumn{2}{c}{Hyper-parameter} & Value  \\ \midrule
\multicolumn{1}{l|}{\multirow{6}{*}{VQ Encoder}}           & \multicolumn{1}{l|}{Encoder Layers}                 & 3    \\
\multicolumn{1}{l|}{}                                            & \multicolumn{1}{l|}{Hidden Size}                    & 128  \\
\multicolumn{1}{l|}{}                                            & \multicolumn{1}{l|}{Conv1D Kernel}                  & 3    \\ 
\multicolumn{1}{l|}{}                                            & \multicolumn{1}{l|}{Conv1D Dilation}                & [1,1,1]    \\ \multicolumn{1}{l|}{}                                            & \multicolumn{1}{l|}{VQ Embedding Size}              & 128    \\ \multicolumn{1}{l|}{}                                            & \multicolumn{1}{l|}{VQ Embedding Channel}           & 32    \\ \midrule
\multicolumn{1}{l|}{\multirow{4}{*}{Masked Decoder}}           & \multicolumn{1}{l|}{Decoder Layers}                 & 3    \\
\multicolumn{1}{l|}{}                                            & \multicolumn{1}{l|}{Hidden Size}                    & 128  \\
\multicolumn{1}{l|}{}                                            & \multicolumn{1}{l|}{Conv1D Kernel}                  & 3    \\ 
\multicolumn{1}{l|}{}                                            & \multicolumn{1}{l|}{Conv1D Dilation}                & [1,2,1]    \\ \midrule
\multicolumn{1}{l|}{\multirow{4}{*}{Localizer}}           & \multicolumn{1}{l|}{Decoder Layers}                 & 3    \\
\multicolumn{1}{l|}{}                                            & \multicolumn{1}{l|}{Hidden Size}                    & 128  \\
\multicolumn{1}{l|}{}                                            & \multicolumn{1}{l|}{Conv1D Kernel}                  & 3    \\ 
\multicolumn{1}{l|}{}                                            & \multicolumn{1}{l|}{Conv1D Dilation}                & [1,2,1]    \\ \midrule
\multicolumn{1}{l|}{\multirow{4}{*}{Restorer}}           & \multicolumn{1}{l|}{Decoder Layers}                 & 3    \\
\multicolumn{1}{l|}{}                                            & \multicolumn{1}{l|}{Hidden Size}                    & 128  \\
\multicolumn{1}{l|}{}                                            & \multicolumn{1}{l|}{Conv1D Kernel}                  & 3    \\ 
\multicolumn{1}{l|}{}                                            & \multicolumn{1}{l|}{Conv1D Dilation}                & [1,2,1]    \\ \midrule
\multicolumn{1}{l|}{\multirow{4}{*}{Manipulator}}      & \multicolumn{1}{l|}{Decoder Layers}                                 & 4    \\
\multicolumn{1}{l|}{}                                            & \multicolumn{1}{l|}{Hidden Size}                 & 128    \\
\multicolumn{1}{l|}{}                                            & 
\multicolumn{1}{l|}{Filter Size}                 & 512    \\
\multicolumn{1}{l|}{}                                            & \multicolumn{1}{l|}{Kernel Size}                 & 5    \\
\multicolumn{1}{l|}{}                                            & \multicolumn{1}{l|}{Code Embedding Size} & 128    \\
\multicolumn{1}{l|}{}                                            & \multicolumn{1}{l|}{Attention Headss} & 2    \\ \midrule
\multicolumn{2}{c|}{Total Number of Parameters}                                                                                                           & 6.22M                                  \\ \bottomrule
\end{tabular}
\end{table}

\subsection{A.3 Training Details}  
We train DiscreteWM on 8 NVIDIA A100 GPU, with a batch size of 20 sentences. We use the Adam optimizer with $\beta_1 = 0.9$, $\beta_2 = 0.999$, $\epsilon = 10^{-9}$ and follow the same learning rate schedule in~\cite{vaswani2017attention}. During training, the mask ratio $\gamma$ is randomly sampled from $\text{Uniform}(0.1, 0.5)$ for each training step. It takes 200k steps for the first stage model's training (the VQ encoder, masked decoder, localizer, and restorer) and 100k steps for the second stage model's training (the manipulator) until convergence. During the first-stage training, the overall loss can be formulated as:
\begin{equation}
\label{eq:eq_7}
\mathcal{L}_{1st} = \mathcal{L}_{loc} + \lambda_{res}\mathcal{L}_{res} + \mathcal{L}_{VQ}
\end{equation} 
\begin{equation}
\label{eq:eq_7_2}
~~\mathcal{L_{VQ}} = \mathcal{L}_{\mathrm{rec}} + \mathcal{L}_{\mathrm{code}} + \lambda_{adv}\mathcal{L}_{\mathrm{adv}},
\end{equation} 
where $\lambda_{res}$ and $\lambda_{adv}$ are hyper-parameters to balance these terms. $\mathcal{L}_{loc}$, $\mathcal{L}_{res}$, and $\mathcal{L}_{VQ}$ represent the training loss of the localizer, restorer, and the VQ-VAE, respectively. $\lambda_{adv}$ is set to $10^{-2}$. In the first 100k steps of the first-stage training, $\lambda_{res}$ is set to $1$ to learn robust encoding and detection capabilities. Then, $\lambda_{res}$ is set to $0.5$ to enhance the imperceptibility.

\subsection{A.4 Random Seeds}
We ran the experiments with 10 different random seeds (1234,1111,2222,3333,4444,5555,6666,7777,8888,9999) and obtained the averaged results.

\subsection{A.5 About the Setting of Baselines}
\label{app:baseline_setting}
\begin{enumerate}
    \item For Audiowmark~\cite{westerfeld2020audiowmark}, we use its default setting (i.e., the strength is set to 10 and the length of the payload is set to the standard type). 
    \item For DeAR~\cite{liu2023dear}, we successfully implement their algorithm and achieve comparable results of their paper.
    \item For Chang Liu's method~\cite{liu2023detecting}, we use the 30 BPS version of their pre-trained model.
    \item For WavMark~\cite{chen2023wavmark}, we use its official implementation and pre-trained parameters.
    \item For Seamless~\cite{chen2023wavmark}, we successfully reproduce their model and achieve comparable results of their paper.
\end{enumerate} 

\subsection{A.6 About the segment-based evaluation and utterance-level evaluation}
We use the segment-based evaluation for information hiding in Section 4.2 and use the utterance-level evaluation for AI-generated content detection in Section 4.3. In the segment-based evaluation, the carrier speech is only one second long, which will greatly increase the difficulty of watermarking. We use this setting to better illustrate the differences between different methods. Besides, DeAR can not directly be applied to the utterance-based scenario. Therefore, we use segment-based evaluation for information hiding. On the other hand, for AI-generated content detection, utterance-level evaluation is more in line with practical application scenarios.

\section{B\quad Details of Distortions}
\label{app:details_distortion}
Due to the limited page space, our experiments in Section 4 consider the following common distortion types:
\begin{enumerate}
    \item \textit{Gaussian Noise (GN)}: a Gaussian noise signal was introduced into the speech, while ensuring a Signal-to-Noise Ratio (SNR) range of 20 $\sim$ 40 dB.
    \item \textit{Amplitude Scaling (AS)}: decreasing the amplitude of the speech signal to 90\% of its original level.
    \item \textit{Re-Sampling (RS)}: Converting the sampling rate to either twice or half of the original, followed by re-conversion to the original frequency.
    \item \textit{MP3 Compression (MP3)}: Converting the speech clip to the MP3 format at 64 kbps and then converting it back. 
    \item \textit{Median Filter (MF)}: Applying a filter kernel size of 3 to smooth the signal. 
    \item \textit{Low-pass Filter (LP)}: Using a low-pass filter with a cutoff frequency of 5 kHz to remove the high-frequency components in the speech. 
    \item \textit{Echo Addition (EA)}: Attenuating the audio volume by a factor of 0.1 $\sim$ 0.3, delaying it by 100 $\sim$ 300 ms, and then overlaying it with the original. 
\end{enumerate}
Additionally, we also evaluate our method under the following distortions. The experimental settings are consistent with the settings in Section 4.2. The results are shown in Table~\ref{table:additional_results_for_information_hiding}. It can be seen that compared to baseline systems, our approach simultaneously achieves state-of-the-art levels in terms of imperceptibility and robustness.
\begin{enumerate}
    \item \textit{Quantization (QTZ)}: Quantizing the sample points to $2^{8}$ levels.
    \item \textit{Sample Suppression (SS)}: Randomly setting 0.1\% of the sample points to zero.
    \item \textit{Pink Noise (PN)}: a type of random noise characterized by having equal energy per octave, meaning that each octave carries an equal amount of energy. The noise amplitude ratio is set to 0.1.
\end{enumerate}

\begin{table*}[h]
\caption{Additional information hiding results under quantization, sample suppression, pink noise, and vocoder reconstruction distortions.}
\label{table:additional_results_for_information_hiding}
\vskip 0.15in
\begin{center}
\begin{small}
\begin{tabular}{l|ccc|cccc}
\toprule
\multirow{2}{*}{Models} & \multirow{2}{*}{BPS($\uparrow$)} & \multirow{2}{*}{PESQ($\uparrow$)} & \multirow{2}{*}{SNR($\uparrow$)} & \multicolumn{3}{c}{BER(\%)($\downarrow$)} \\
    & & &        & QTZ   & SS   & PN   \\
\midrule
Chang Liu's         & 30 & 3.97 & 24.18 & 0.05 & 0.01 & 0.07 \\
WavMark             & 32 & 4.31 & 38.61 & 3.51 & 1.22 & 1.30 \\
\midrule
Ours-\textit{32bps} & 32 & 4.45 & 38.08 & 0.09 & 0.09 & 0.12 \\
\bottomrule
\end{tabular}
\end{small}
\end{center}
\vskip -0.1in
\end{table*}

\begin{table}[h]
\caption{Results for random mask selection on the information hiding task. ``MEAN'' represents the average BER across all distortions. Ours-\textit{random}-\textit{50} denotes we randomly select the watermark positions for 50 times.}
\label{table:random_mask_selection}
\vskip 0.15in
\begin{center}
\begin{tabular}{l|cc|c}
\toprule
Setting & PESQ($\uparrow$) & SNR($\uparrow$) & MEAN($\downarrow$) \\
\midrule
Ours-\textit{random}-\textit{1}  & 4.46 & 37.93 & 0.29 \\
Ours-\textit{random}-\textit{10} & 4.46 & 38.64 & 0.27 \\
Ours-\textit{random}-\textit{50} & 4.48 & 39.70 & 0.27 \\
\bottomrule
\end{tabular}
\end{center}
\vskip -0.1in
\end{table}

\begin{table}[h]
\caption{Comparisons for different input types. ``MEAN'' represents the average BER across all distortions.}
\label{table:diferent_input_types}
\vskip 0.15in
\begin{center}
\begin{tabular}{l|cc|c|c}
\toprule
Setting & PESQ($\uparrow$) & SNR($\uparrow$) & MEAN($\downarrow$) & RTF($\downarrow$) \\
\midrule
spectrogram  & 4.37 & 38.01 & 0.32 & 0.0044 \\
wave & 4.31 & 33.59 & 0.53 &  0.0039 \\
\bottomrule
\end{tabular}
\end{center}
\vskip -0.1in
\end{table}

\section{C\quad Random mask selection.}
\label{app:random_mask_selection}
Since our method is frame-wise, we can iteratively select the frames where the watermarks are embedded to further improve the imperceptibility. However, currently, an efficient algorithm for selecting the watermarked frames with the highest imperceptibility is lacking. Additionally, the frame-by-frame recursive searching is excessively time-consuming. Therefore, we choose to randomly select watermark positions repeatedly and use the set of watermark positions that offer the best imperceptibility. Due to the high computational cost, we do not use the entire test set in previous experiments. Instead, we randomly selected 2,000 audio samples from the 24,837 test samples to construct the new test set. We set the encoding capacity of all systems to 32 BPS. The results are in Table~\ref{table:random_mask_selection}. It can be seen that the imperceptibility (SNR) of our method can be further improved by the mask selection techniques. However, since the variances of PESQ and BER are relatively small, the mask selection mechanism has minor improvements for them.

\section{D \quad Inference Strategies of DiscreteWM}
Below is a detailed schematic representation of the algorithmic process.
\label{app:suanfatu}

\begin{algorithm}[h]
\caption{Inference Strategies of DiscreteWM}
\label{algorithm:inference_strategies}
\begin{algorithmic}
\STATE \textbf{Input:} clean speech, y \\ 
\hspace{10.5mm}watermark string, $w$ \\
\IF{\textit{Information hiding}}
\STATE
\begin{enumerate}
\item Transform $y$ to discrete tokens $z$ and apply the manipulater model $\mathbf{M}$ to get $P(z^{(t)}_{k})$ for each watermarked frame $t$.
\item Sample the watermarked tokens from $P(z^{(t)}_{k})$ and make sure that the sampled tokens have the same modular arithmetic relation with the embedded watermark string.
\item Reconstruct the watermarked speech $\hat{y}$ and decode the watermarks from $\hat{y}$ with $\mathbf{D}$ and $\mathbf{R}$
\end{enumerate}
\ELSIF{\textit{AI-generated content detection}}
\STATE
\begin{enumerate}
\item Reconstruct a portion of frames of $y$ to produce $\hat{y}$
\item Use $\mathbf{D}$ to obtain the number of watermarked frames and calculate the Z-statistic
\item Detect the utterance-level watermark when the Z-statistic is larger than a pre-defined threshold
\end{enumerate}
\ENDIF
\end{algorithmic}
\end{algorithm}

\section{E \quad Spectrogram VQ vs Wave VQ}
\label{app:spec_vs_wav}
In previous works, some of them utilize spectrogram features~\cite{chen2023wavmark,liu2023detecting} while others directly use waveform as input~\cite{seamless_communication}. Different input types will affect the overall performance, inference speed, and other metrics of the model. Therefore, this section mainly discusses whether to use Spectrogram VQ or Wave VQ for DiscreteWM. We set the encoding capacity of all systems to 32 BPS. The results of the AI-generated speech detection task are shown in Table~\ref{table:diferent_input_types}. ``Ours-spectrogram'' is the original version of DiscreteWM. ``Ours-wave'' adopts the backbone architecture of Encodec~\cite{defossez2022high} so that it can take waveform as inputs. We keep the vector quantization module and total parameters of the model consistent between the two systems. In terms of inference speed, both systems are very efficient. Although the STFT and iSTFT process is relatively time-consuming, the waveform encoder also requires a down-sampling process and more complicated network architecture. In terms of imperceptibility and robustness, the performance of the spectrogram-based system is slightly better. Compared to the waveform, the magnitude spectrogram is easier for the model to spectrogram. Besides, we concatenate the ground-truth phase spectrum to the output of ``Ours-spectrogram''. Compared to ``Ours-spectrogram'', ``Ours-wave'' has to learn the complicated distribution of phase spectrogram.

\section{F \quad Disccusions about the Vector Quantised Discrete Representation}
\label{app:disccusions_about_vqvae_jsp}
Vector-quantized variational autoencoder (VQVAE) is a method that learns to discretize continuous features into discrete space using a limited number of codebook vectors~\cite{van2017neural,zheng2023online}. This discrete feature is typically used as an intermediate representation for downstream generation tasks, such as image generation~\cite{esser2021taming,razavi2019generating,hu2022global}, video generation~\cite{rakhimov2020latent,yan2021videogpt}, and speech synthesis~\cite{du2022vqtts,wang2023neural,yang2023instructtts,shen2023naturalspeech}. In the field of speech synthesis, VQVAE can compress speech information into a more compact latent space. Compared to directly using continuous waveform or mel-spectrogram as the training target, using latent discrete codes as intermediate features can reduce the difficulty of learning and improve the overall performance of speech synthesis models~\cite{shen2023naturalspeech}. Inspired by these, we construct a robust discrete latent space and integrate the robust discrete intermediate representation into the speech watermarking framework to ensure the robustness of our scheme.

\section{G \quad Disccusions about different architectures}

we conduct experiments for the ours-32bps setting with different manipulator architectures, including Conv1d (similar to the architecture of the masked decoder used in our model) and Conformer. As shown in the following Table \ref{table:diferent_arch_types}, attention-based models such as Transformer and Conformer perform well, while purely convolutional structures show relatively lower performance. 

\begin{table}[h]
\caption{Comparisons for different manipulator architectures. ``MEAN'' represents the average BER across all distortions.}
\label{table:diferent_arch_types}
\vskip 0.15in
\begin{center}
\begin{adjustbox}{width=0.48\textwidth}
\begin{tabular}{l|cc|c}
\toprule
Manipulator Architectures & PESQ($\uparrow$) & SNR($\uparrow$) & MEAN($\downarrow$) \\
\midrule
Conv1d &	4.305 &	36.97 &	0.54\\
Transformer	& 4.451 &	38.08 &	0.28\\
Conformer &	4.453 &	38.14 & 0.27\\
\bottomrule
\end{tabular}
\end{adjustbox}
\end{center}
\vskip -0.1in
\end{table}

\section{H \quad Disccusions about different codebook lengths}
An increased codebook length would enhance audio quality and encoding capacity. However, once the codebook length reaches a certain threshold, further increases do not yield additional performance gains, which is similar to the experimental results in the field of speech synthesis~\cite{ji2024wavtokenizer,ji2024wavchat}. The corresponding experimental results are shown in Table~\ref{table:diferent_code_length}. We observed that the model reaches its best performance once the number of discrete tokens exceeds 128. Therefore, as outlined in Appendix A.1, we use a VQ codebook with 128 codes and an embedding size of 128.

\begin{table}[h]
\caption{Comparisons for different codebook length. ``MEAN'' represents the average BER across all distortions.}
\label{table:diferent_code_length}
\vskip 0.15in
\begin{center}
\begin{adjustbox}{width=0.48\textwidth}
\begin{tabular}{l|cc|c}
\toprule
Codebook length & PESQ($\uparrow$) & SNR($\uparrow$) & MEAN($\downarrow$) \\
\midrule
24	&3.63	&26.55	&0.25\\
64	&4.18	&32.30	&0.29\\
128	&4.45	&38.08	&0.28\\
256	&4.42	&38.11	&0.36\\

\bottomrule
\end{tabular}
\end{adjustbox}
\end{center}
\vskip -0.1in
\end{table}

\section{I \quad Limitations and Future Work}
In this section, we discuss the limitations of the proposed method and outline our plans for future work to address them. Firstly, although the manipulator model helps us to select the watermarked code, different codes in the discrete codebook have different characteristics (e.g., robustness and imperceptibility). Our method lacks an appropriate way to analyze the characteristics of codes. We plan to address this problem by designing further experiments and visualizations for the codebook vector. Secondly, in this paper, we adopt a GAN-based architecture for efficient speech watermarking. However, the diffusion-based models have shown superior performance on various tasks. We will investigate the application of diffusion-based models for speech watermarking. Finally, the inference speed can be further improved by introducing more efficient network structures.

\section{J \quad Impact Statements}
In the era of large-scale voice models, AI security and privacy preservation are particularly important. Speech watermarking technology offers a proactive and efficient solution for copyright protection, voice source tracking, and defense against voice cloning attacks. Our DiscreteWM enhances the overall robustness and imperceptibility and addresses the fixed length issues for speech watermarking. With its versatility and flexibility, our technology will enhance security and trust in voice-based applications, thereby facilitating individual users, social media, and cloud service providers. Generally speaking, our scheme will not raise ethical concerns in society. On the contrary, our approach will restrict the development of voice spoofing and guarantee the security of online voice services.

\end{document}